\documentclass[sigconf]{acmart}
\usepackage{multirow}
\usepackage{booktabs}
\usepackage{enumitem}
\usepackage{float}

\AtBeginDocument{%
  \providecommand\BibTeX{{%
    \normalfont B\kern-0.5em{\scshape i\kern-0.25em b}\kern-0.8em\TeX}}}

\graphicspath{{./images/}} 
\begin{document}

\title{RAG-Enhanced Large Language Models for Dynamic Content Expiration Prediction in Web Search}

\author{Tingyu Chen}
\affiliation{%
  \institution{Baidu Inc.}
  \city{Beijing}
  \country{China}
}
\email{tingyuchen.cty@gmail.com}

\author{Wenkai Zhang}
\affiliation{%
  \institution{Baidu Inc.}
  \city{Beijing}
  \country{China}
}
\email{zhangwk0106@gmail.com}

\author{Li Gao}
\authornote{Corresponding author.}
\affiliation{
  \institution{Baidu Inc.}
  \city{Beijing}
  \country{China}
}
\email{gaoli05@baidu.com}

\author{Lixin Su}
\affiliation{
  \institution{Baidu Inc.}
  \city{Beijing}
  \country{China}
}
\email{sulixinict@gmail.com}

\author{Ge Chen}
\affiliation{
  \institution{Baidu Inc.}
  \city{Beijing}
  \country{China}
}
\email{chenge02@baidu.com}

\author{Dawei Yin}
\affiliation{
  \institution{Baidu Inc.}
  \city{Beijing}
  \country{China}
}
\email{yindawei@acm.org}

\author{Daiting Shi}
\affiliation{
  \institution{Baidu Inc.}
  \city{Beijing}
  \country{China}
}
\email{shidaiting01@baidu.com}

\renewcommand{\shortauthors}{Tingyu Chen et al.}

\begin{abstract}
  In commercial web search, aligning content freshness with user intent remains challenging due to the highly varied lifespans of information. Traditional industrial approaches rely on static time-window filtering, resulting in "one-size-fits-all" rankings where content may be chronologically recent but semantically expired. To address the limitation, we present a novel Large Language Models (LLMs)-based Query-Aware Dynamic Content Expiration Prediction Framework deployed in Baidu search, reformulating timeliness as a dynamic validity inference task. Our framework extracts fine-grained temporal contexts from documents and leverages LLMs to deduce a query-specific "validity horizon"—a semantic boundary defining when information becomes obsolete based on user intent. Integrated with robust hallucination mitigation strategies to ensure reliability, our approach has been evaluated through offline and online A/B testing on live production traffic. Results demonstrate significant improvements in search freshness and user experience metrics, validating the effectiveness of LLM-driven reasoning for solving semantic expiration at an industrial scale.
\end{abstract}




\maketitle

\section{Introduction}
Modern search engines must align semantic relevance with timeliness \cite{llmirsurvey1}. Unlike static knowledge retrieval, assessing \textit{temporal relevance} to determine whether content remains valid or has become stale requires dynamically reasoning about information lifespans based on user intent 
\cite{llmtemporalreasoning1, llmtemporalreasoning2,llmtemporalreasoning3,llmragtuprompt}.

Traditionally, industrial search engines rely on rule-based bucketing to manage timeliness \cite{oldtime1, oldtime2, oldtime3}. A common practice is to apply a global "recency boost" to documents published within a fixed window (e.g., the past 30 days) or to indiscriminately penalize older content. While computationally efficient, this "one-size-fits-all" strategy suffers from a severe granularity mismatch.

\begin{figure}[htb]
  \centering
  \includegraphics[width=\linewidth]{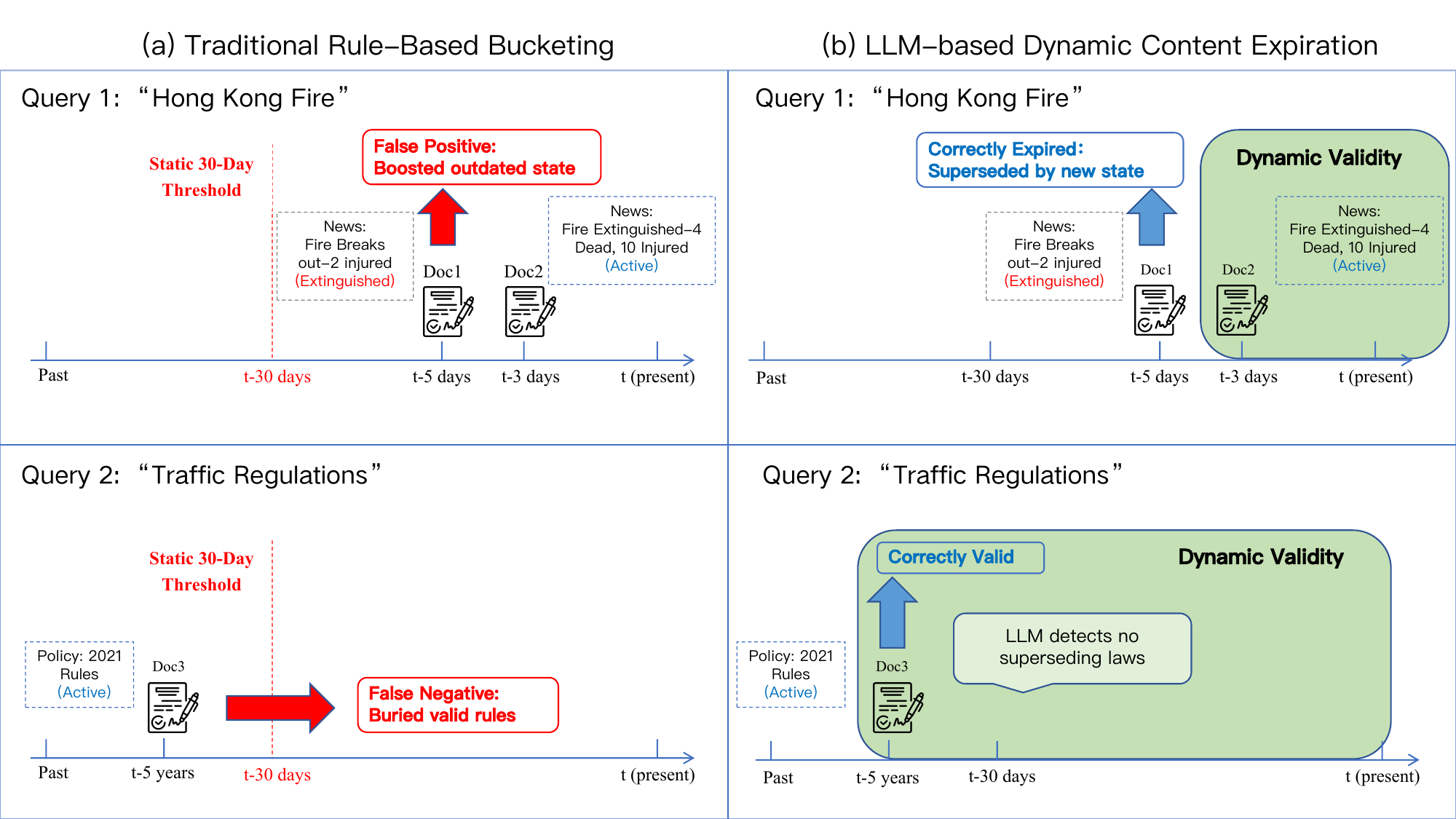}
  \caption{Comparison between (a) traditional rule-based bucketing and (b) our proposed LLM-based dynamic content expiration prediction framework. By inferring a query-specific validity horizon, our approach correctly expires superseded news while preserving long-term authoritative content.}
  \label{fig:comparison}
\end{figure}

As illustrated in Figure \ref{fig:comparison}, a universal expiration threshold does not exist. For a breaking news query such as "Hong Kong Fire", a report published just two days ago may already be expired if the fire has been extinguished and the situation has evolved. Conversely, for a query like "Traffic Regulations", a policy document published several years ago remains entirely valid. Fundamentally, these limitations stem from treating timeliness as a static metadata attribute rather than a dynamic semantic relationship between query intent and content timelines.

To bridge this gap, we propose a novel LLM-based dynamic content expiration prediction framework. Building upon the Retrieval-Augmented Generation (RAG) paradigm \cite{RAGllm1, timerag, suffcontextrag}, we reformulate the timeliness challenge as a dynamic content expiration prediction problem. Specifically, we first introduce a temporal information extraction module that segments raw document content ($\mathcal{D}$) into time-aware focused chunks ($S_{focus}$). This allows the model to prioritize salient temporal signals and mitigate noise from long-form text \cite{llmprompttimereason,longcontext,longcontext-qa}. Building on this, we employ a structured prompting mechanism for temporal inference. The LLM \cite{tempretriever,llmtemporal1} synthesizes the query intent ($Q$) and the extracted temporal context to dynamically determine a query-specific expiration threshold ($t_{exp}^*$), effectively decoupling validity from rigid rule-based thresholds. To counter the hallucination risks inherent in generative models \cite{halluci1}, we incorporate a temporal consistency verification module using a Contrastive Forward-Backward Chain-of-Thought (CoT) \cite{cot_origin,fb-cot,fb-cot1} mechanism. This self-correcting process validates the reasoning path to ensure high confidence. Finally, to meet strict online latency constraints, we transform these LLM-derived semantic judgments into lightweight binary signals ($0/1$ flags) by directly comparing document timeliness factors against the inferred threshold, which are then seamlessly integrated into Baidu's core ranking model. In summary, our contributions are as follows:

\begin{itemize}[topsep=0pt, partopsep=0pt, parsep=0pt, leftmargin=*]
    \item We propose a scalable LLM-based query-aware dynamic content expiration prediction framework that replaces traditional rule-based bucketing with fine-grained temporal validity deduction, addressing the timestamp-validity mismatch in industrial search.
    \item We design a robust hallucination suppression mechanism tailored for temporal reasoning, which integrates contrastive forward-backward CoT with source authority signals to ensure high precision in identifying expired content.
    \item Our approach has been successfully deployed online to power Baidu Search, delivering positive gains in search engagement.
\end{itemize}

\section{Method}
\begin{figure*}[htbp] 
  \centering
  \includegraphics[width=1\textwidth]{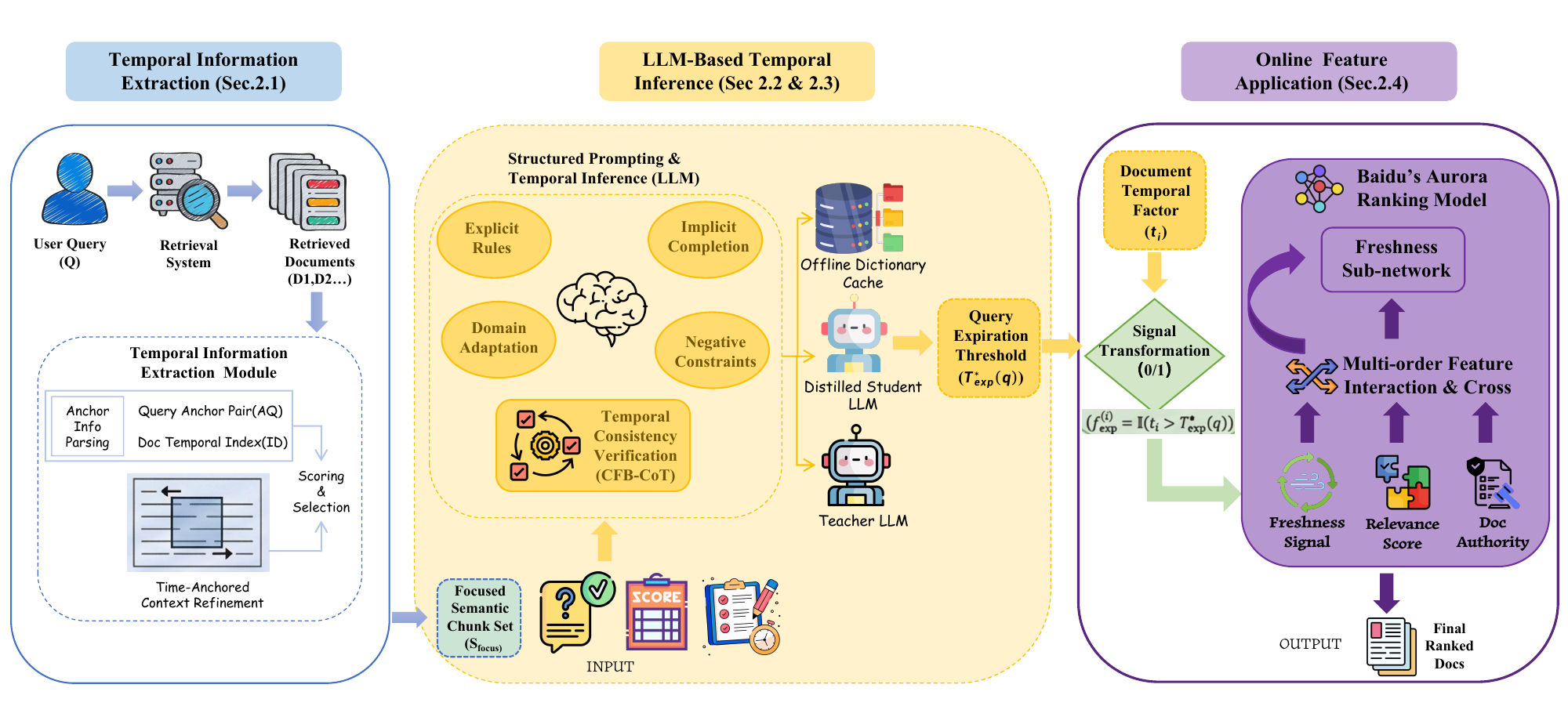} 
  \caption{An overview of our proposed LLM-based Query-Aware Dynamic Content Expiration Prediction Framework.}
  \label{fig:framework}
\end{figure*}

In this section, we introduce our proposed framework as Figure~\ref{fig:framework} shows.

\subsection{Temporal Information Extraction Module}
\label{sec:extraction}
Acting as a post-retrieval refinement mechanism, this module adapts "document-time anchoring" \cite{temporalinfoextract} into a discrete text-level framework. This module distills a raw document $\mathcal{D}$ into a concise \textit{Focused Semantic Chunk Set} $S_{focus}$ to mitigate noise and attention dispersion during LLM reasoning.

\subsubsection{Anchor Information Parsing}
Given a query $Q$ and candidate documents $\mathcal{D}$, we first extract a query anchor pair $\mathcal{A}_Q = (K_Q, T_Q)$, comprising semantic keywords $K_Q$ and explicit or implicit temporal entities $T_Q$ (where $T_Q = \emptyset$ if undetected). Simultaneously, a temporal parser traverses $\mathcal{D}$ to construct a Document Temporal Index $\mathcal{I}_D$, directly linking temporal expressions $t^D_j$ to sentence indices via a $\{Time \rightarrow Position\}$ mapping.

\subsubsection{Time-Anchored Context Refinement}
Using $\mathcal{I}_D$, we define a sliding window of size $L$ (default $L=5$ sentences) centered on each anchor sentence, yielding an initial candidate chunk set $S_{candidate}$. If $\mathcal{I}_D = \emptyset$, we extract the document title and leading paragraphs as a fallback. Next, to identify the most relevant context, we calculate a composite score $S_{rel}(c)$ for each text chunk $c \in S_{candidate}$:
\begin{equation}
S_{rel}(c) = \mathbb{I}_{K}(c \cap K_Q \neq \emptyset) \cdot \left[ \alpha \cdot Rel_K(c, K_Q) + (1-\alpha) \cdot Rel_T(c, T_Q) \right]
\end{equation}
where $\mathbb{I}_{K}$ is an indicator function equaling 1 if the chunk $c$ contains any query keywords and 0 otherwise. The hyperparameter $\alpha=0.6$ balances semantic similarity ($Rel_K$, via cosine similarity) and temporal alignment ($Rel_T$). Specifically, $Rel_T$ prioritizes exact matches and hierarchical overlaps (e.g., '2025' vs. '2025 Q3'). Crucially, if the query lacks explicit temporal constraints ($T_Q = \emptyset$), $Rel_T$ dynamically shifts to an exponential recency decay function $\exp(-\lambda \Delta t)$, where $\Delta t$ denotes the elapsed time. Chunks exceeding a threshold $\tau$ form the final \textit{Focused Semantic Chunk Set} $S_{focus}$, serving as the direct input for subsequent LLM reasoning.

\subsection{Structured Prompting for Temporal Inference}
\label{sec:prompting}

\subsubsection{Core Prompting Principles}
We design a structured prompt framework to guide the LLM to perform precise temporal reasoning over $S_{focus}$. Specifically, the prompt integrates the user query, $S_{focus}$, and the search time, alongside other contextual metadata, to establish a strict absolute time anchor. The prompt establishes a logical blueprint for expiration inference (e.g., inferring a target time from start times and validity periods) and instructs the model to leverage common sense for implicit date completion. Furthermore, we integrate domain adaptation via few-shot prompting to recognize periodic events and ensure robustness across diverse query types. Finally, we employ negative constraints and CoT reasoning to prevent the model from extracting irrelevant auxiliary dates or conflating distinct temporal entities. The specific mechanisms for this consistency verification are detailed in Section \ref{sec:consistency}.

\subsubsection{Temporal Optimization Objective}
To enhance temporal precision, we define a joint optimization objective $\mathcal{L}_{temp}$ via a reward-based alignment framework. Functioning as a negative reward signal, the alignment penalty is formulated as:
\begin{equation}
\mathcal{L}_{temp} = \mathcal{D}_{time}(t_{init}, t_{gt}) + \lambda_1 \cdot \mathcal{L}_{gran} + \lambda_2 \cdot \mathcal{L}_{cons}
\label{eq:temp_loss}
\end{equation}
where $t_{init}$ and $t_{gt}$ represent the LLM's initial predicted candidate and the ground-truth timestamp, respectively. Here, $\mathcal{D}_{time}(\cdot)$ measures the normalized temporal distance. $\mathcal{L}_{gran}$ penalizes temporal granularity mismatch, while $\mathcal{L}_{cons}$ penalizes logical contradictions during CoT reasoning (both formally defined in Section \ref{sec:consistency}). Hyperparameters $\lambda_1$ and $\lambda_2$ are empirically tuned to balance these objectives.

\subsection{Temporal Consistency Verification Module}
\label{sec:consistency}
To resolve conflicting temporal cues in $S_{focus}$ and mitigate LLM hallucinations, we implement a dual-verification architecture. 

The first mechanism, \textbf{Consistency Forward-Backward Reasoning}, acts as an internal check. The LLM generates an initial candidate $t_{init}$ via forward inference, then performs backward stress-testing against the retrieved evidence to verify logical coherence. We quantify this alignment using a consistency penalty: $\mathcal{L}_{cons} = 1 - S_{self}(\text{traj}_{fwd}, \text{traj}_{bwd})$, where $S_{self}$ is the LLM's self-evaluated logical consistency score to capture conflicts. To penalize precision inconsistencies, we introduce a granularity penalty:
\begin{equation}
\mathcal{L}_{gran} = | D(t_{init}) - D(t_{gt}) |
\end{equation}
where $D(\cdot)$ maps the temporal resolution to a hierarchical depth level (e.g., year=1, month=2, day=3), penalizing the hierarchical distance of mismatched tiers. Minimizing these non-differentiable penalties via reward-based alignment ensures coherent logical scales and effectively reduces hallucinated drifts.

The second mechanism, \textbf{Authority-Weighted Evidence Fusion}, resolves external conflicts among divergent sources by maximizing collective support for the optimal expiration timestamp $t_{exp}^*$:
\begin{equation}
t_{exp}^* = \arg\max_{t_c \in \mathcal{T}_c} \sum_{c \in S_{focus}} w_c \cdot \mathbb{I}_{align}(t_c, c)
\end{equation}
where $\mathcal{T}_c$ denotes the candidate timestamps generated by the LLM. Here, $w_c = Auth(c) \cdot S_{rel}(c)$ serves as the joint chunk weight, combining source authority and semantic relevance (Section \ref{sec:extraction}). The indicator $\mathbb{I}_{align}(t_c, c)$ equals 1 if $t_c$ logically aligns with evidence $c$. This robust mechanism ensures the most authoritative evidence dictates the output.

\subsection{Online Feature Application}
\label{sec:online}

\subsubsection{Signal Transformation and Parameter Acquisition}
For any query-document pair $(q, d_i)$, we compare the query's expiration threshold $t_{exp}^*(q)$ with the document's temporal factor $t_i$ to generate a binary prior signal:

\begin{equation}
f_{\text{exp}}^{(i)} = \mathbb{I}(t_i > t_{exp}^*(q))
\end{equation}

where $\mathbb{I}(\cdot)$ is the indicator function. A value of 1 denotes the document is LLM-validated as fresh, while 0 indicates expiration. 
For the query threshold $t_{exp}^*(q)$ is obtained from an offline dictionary cache, a distilled student LLM, or via direct online calls to the teacher LLM. Regarding the document temporal factor $t_i$, the publication time ($t_{pub}$) can be used as it typically correlates with the occurrence of events in most scenarios. Alternatively, an intrinsic content time can be extracted via LLM-based semantic comprehension.

\subsubsection{Robustness and Circuit Breaking}
To ensure system reliability, we implement a \textit{Circuit Breaking \& Fallback} mechanism. If real-time inference times out, the generated threshold exhibits significant distributional shifts, or cache misses occur, the system triggers a failsafe mode where $f_{\text{exp}}^{(i)}$ defaults to 0, gracefully degrading to the baseline ranking without introducing noise. 

\subsubsection{Deep Ranking Model Integration}
Finally, $f_{\text{exp}}^{(i)}$ is injected into Baidu's ranking architecture as a learnable soft constraint via multi-order feature interactions. Specifically, within the Freshness sub-network, $f_{\text{exp}}^{(i)}$ interacts with other core signals:
\begin{equation}
\text{Score}(d_i, q) = \Phi \left( \mathbf{x}_i, f_{\text{exp}}^{(i)}, \text{Cross}(f_{\text{exp}}^{(i)}, S_{rel}(d_i, q), Auth(d_i), \dots); \Theta \right)
\end{equation}
This formulation converts the discrete temporal prior into adaptive weights, enabling Aurora to dynamically balance freshness against other factors for a globally optimal ranking.

\section{Experiments}
To verify the effectiveness of the proposed \textit{LLM-based dynamic content expiration prediction framework}, we conducted three progressive experiments: offline model evaluation, online A/B testing on Baidu Search, and human evaluation. All experiments compare the production \textit{Aurora} ranking model (denoted as \textbf{Aurora-Baseline}) against the experimental model fused with the binary expiry time feature $f_{\text{exp}}$ (denoted as \textbf{Aurora-Expiry}). Statistical significance was verified for all results ($p<0.05$).

\subsection{Offline Model Evaluation}

The dataset was constructed from Baidu Search logs over a three-month period, comprising 1 million valid queries and 10 million candidate documents across diverse domains. 

Table~\ref{tab:offline_results} shows that Aurora-Expiry significantly enhances overall search quality, evidenced by an increased global Satisfaction Score and a substantial PNR surge for High-Freshness queries. Crucially, the simultaneous decrease in both \textit{day\_away@4} and \textit{day\_away@10} confirms a synchronized freshness improvement across the ranking list, demonstrating that the LLM-derived $f_{\text{exp}}$ acts as an effective semantic filter to precisely penalize obsolete content. Unlike naive recency boosts that blindly promote new but less relevant documents, the concurrent uplift in the Satisfaction Score proves that our approach improves timeliness without compromising semantic accuracy. Furthermore, other general relevance metrics across the random test set remained highly stable with negligible fluctuations, confirming the framework's safety and robustness in large-scale production.

\begin{table}[H]
\centering
\caption{Offline Evaluation Results. Satisfaction Score and Positive Navigation Rate (PNR) measure freshness preference (higher is better). The day\_away metric gauges the median time factor of top results (lower is fresher).}
\label{tab:offline_results}
\begin{tabular}{l c}
\toprule
\textbf{Metric} & \textbf{Rel. Imp. ($\Delta\%$)} \\
\midrule
Satisfaction Score & +0.52\% \\
PNR (Tier 2: $\le$ 1 month) & +5.21\% \\
PNR (Tier 3: $\le$ 1 week) & +7.36\% \\
\midrule
day\_away@4 & -2.44\% \\
day\_away@10 & -3.19\% \\
\bottomrule
\end{tabular}
\end{table}

\subsection{Online A/B Test}
To evaluate the real-world impact, we conducted a 14-day A/B test on Baidu Search's live production traffic (5\% per group) across all platforms. 

As detailed in Table~\ref{tab:online_results}, Aurora-Expiry significantly improves ranking freshness, highlighted by a 12.81\% drop in the median \textit{day\_away@4} for High-Freshness queries.  This consistent reduction in content age across both top-4 and top-10 positions confirms that $f_{\text{exp}}$ effectively filters expired content across the entire ranking list. Crucially, this successful temporal optimization directly drives core business value, yielding significant and consistent gains in both Click-Through Rate (CTR) and Satisfactory Consumption (SC). Coupled with a positive Search Retention Rate, these results prove that the framework goes beyond simply surfacing recent documents to attract clicks; instead, it delivers genuine post-click value by prioritizing valid, up-to-date information, thereby comprehensively elevating the overall search experience.

\begin{table}[H]
\centering
\small
\caption{Online A/B Test Results (Relative Improvement $\Delta\%$).}
\label{tab:online_results}
\setlength{\tabcolsep}{8pt}
\begin{tabular}{l l c c}
\toprule
\textbf{Category} & \textbf{Metric} & \textbf{@4} & \textbf{@10} \\
\midrule
\multirow{4}{*}{User Experience} & Satisfactory Cons. & \multicolumn{2}{c}{+0.78\%} \\
 & CTR & \multicolumn{2}{c}{+0.41\%} \\
 & Search Activity & \multicolumn{2}{c}{+0.12\%} \\
 & Search Retention Rate & \multicolumn{2}{c}{+0.06\%} \\
\midrule
\multirow{2}{*}{Overall Freshness} & day\_away (Median) & -6.87\% & -6.24\% \\
 & day\_away (Mean) & -6.70\% & -5.94\% \\
\midrule
\multirow{2}{*}{High-Freshness} & day\_away (Median) & -12.81\% & -11.46\% \\
 & day\_away (Mean) & -12.94\% & -10.91\% \\
\bottomrule
\end{tabular}
\end{table}

\subsection{Human Evaluation}
To authentically simulate real user perception, we conducted a blind GSB (Good/Same/Bad) human evaluation. As shown in Table~\ref{tab:human_eval}, Aurora-Expiry demonstrates pronounced superiority in complex search contexts. Notably, the ``Long-Tail / Cold Demand'' cohort achieved an outstanding 12:2 Good-to-Bad (G:B) ratio. Conversely, the ``Random Demand'' cohort exhibits a more modest advantage ratio, though still maintaining a solid 5:2 G:B ratio. For such broad queries lacking explicit temporal constraints, this net-positive result confirms that $f_{\text{exp}}$ safely processes general, non-time-sensitive traffic without degrading overall search stability. Furthermore, the strong positive results in the time-sensitive cohort (6:1 G:B ratio) validate the framework's precision in capturing dynamic temporal intents.

\begin{table}[H]
\centering
\caption{Human Evaluation (GSB) across key scenarios.}
\label{tab:human_eval}
\begin{tabular}{l c c}
\toprule
\textbf{Scenario} & \textbf{Advantage Ratio} & \textbf{G:B Ratio} \\
\midrule
Long-Tail / Cold Demand & 5.00\% & 12:2 \\
Time-Sensitive & 2.50\% & 6:1 \\
Random Demand & 1.50\% & 5:2 \\
\bottomrule
\end{tabular}
\end{table}

\section{Conclusion}
In this paper, we propose a novel LLM-based Dynamic Content Expiration framework that shifts the paradigm of timeliness ranking in Baidu Search. Moving beyond rigid rules, our approach extracts fine-grained temporal contexts and applies dual-stage consistency verification to deduce reliable, query-specific validity horizons. By distilling these semantic judgments into lightweight discrete signals $f_{\text{exp}}$, we successfully bridge complex temporal reasoning with the strict latency constraints of a production ranker. Extensive offline evaluations and online A/B test validate the framework’s profound impact on holistic search quality.


\bibliographystyle{ACM-Reference-Format}
\nocite{*} 
\balance
\bibliography{references.bib}

@inproceedings{llmprompttimereason,
    title = "Enhancing Temporal Sensitivity and Reasoning for Time-Sensitive Question Answering",
    author = "Yang, Wanqi  and
      Li, Yanda  and
      Fang, Meng  and
      Chen, Ling",
    editor = "Al-Onaizan, Yaser  and
      Bansal, Mohit  and
      Chen, Yun-Nung",
    booktitle = "Findings of the Association for Computational Linguistics: EMNLP 2024",
    month = nov,
    year = "2024",
    address = "Miami, Florida, USA",
    publisher = "Association for Computational Linguistics",
    url = "https://aclanthology.org/2024.findings-emnlp.848/",
    doi = "10.18653/v1/2024.findings-emnlp.848",
    pages = "14495--14508"
}

@inproceedings{temporalinfoextract,
author = {Song, Xintong and Liang, Bin and Sun, Yang and Zhang, Chenhua and Wang, Bingbing and Xu, Ruifeng},
title = {Bridging Time Gaps: Temporal Logic Relations for Enhancing Temporal Reasoning in Large Language Models},
year = {2025},
isbn = {9798400715921},
publisher = {Association for Computing Machinery},
address = {New York, NY, USA},
url = {https://doi.org/10.1145/3726302.3730173},
doi = {10.1145/3726302.3730173},
booktitle = {Proceedings of the 48th International ACM SIGIR Conference on Research and Development in Information Retrieval},
pages = {3040–3044},
numpages = {5},
location = {Padua, Italy},
series = {SIGIR '25}
}

@misc{llmtemporal1,
      title={ECONET: Effective Continual Pretraining of Language Models for Event Temporal Reasoning}, 
      author={Rujun Han and Xiang Ren and Nanyun Peng},
      year={2021},
      eprint={2012.15283},
      archivePrefix={arXiv},
      primaryClass={cs.CL},
      url={https://arxiv.org/abs/2012.15283}, 
}

@inproceedings{longcontext-qa,
  title={Open temporal relation extraction for question answering},
  author={Shang, Chao and Qi, Peng and Wang, Guangtao and Huang, Jing and Wu, Youzheng and Zhou, Bowen},
  booktitle={3rd Conference on Automated Knowledge Base Construction},
  year={2021}
}

@misc{suffcontextrag,
      title={Sufficient Context: A New Lens on Retrieval Augmented Generation Systems}, 
      author={Hailey Joren and Jianyi Zhang and Chun-Sung Ferng and Da-Cheng Juan and Ankur Taly and Cyrus Rashtchian},
      year={2025},
      eprint={2411.06037},
      archivePrefix={arXiv},
      primaryClass={cs.CL},
      url={https://arxiv.org/abs/2411.06037}, 
}

@misc{RAGllm1,
      title={Emulating Retrieval Augmented Generation via Prompt Engineering for Enhanced Long Context Comprehension in LLMs}, 
      author={Joon Park and Kyohei Atarashi and Koh Takeuchi and Hisashi Kashima},
      year={2025},
      eprint={2502.12462},
      archivePrefix={arXiv},
      primaryClass={cs.CL},
      url={https://arxiv.org/abs/2502.12462}, 
}

@misc{llmtemporalreasoning1,
      title={Timo: Towards Better Temporal Reasoning for Language Models}, 
      author={Zhaochen Su and Jun Zhang and Tong Zhu and Xiaoye Qu and Juntao Li and Min Zhang and Yu Cheng},
      year={2024},
      eprint={2406.14192},
      archivePrefix={arXiv},
      primaryClass={cs.CL},
      url={https://arxiv.org/abs/2406.14192}, 
}

@misc{llmtemporalreasoning2,
      title={Living in the Moment: Can Large Language Models Grasp Co-Temporal Reasoning?}, 
      author={Zhaochen Su and Juntao Li and Jun Zhang and Tong Zhu and Xiaoye Qu and Pan Zhou and Yan Bowen and Yu Cheng and Min zhang},
      year={2024},
      eprint={2406.09072},
      archivePrefix={arXiv},
      primaryClass={cs.CL},
      url={https://arxiv.org/abs/2406.09072}, 
}

@misc{llmtemporalreasoning3,
      title={Large Language Models Can Learn Temporal Reasoning}, 
      author={Siheng Xiong and Ali Payani and Ramana Kompella and Faramarz Fekri},
      year={2024},
      eprint={2401.06853},
      archivePrefix={arXiv},
      primaryClass={cs.CL},
      url={https://arxiv.org/abs/2401.06853}, 
}

@misc{tempretriever,
      title={TempRetriever: Fusion-based Temporal Dense Passage Retrieval for Time-Sensitive Questions}, 
      author={Abdelrahman Abdallah and Bhawna Piryani and Jonas Wallat and Avishek Anand and Adam Jatowt},
      year={2025},
      eprint={2502.21024},
      archivePrefix={arXiv},
      primaryClass={cs.IR},
      url={https://arxiv.org/abs/2502.21024}, 
}

@inproceedings{timerag,
author = {Wang, Zhao and Zhao, Ziliang and Dou, Zhicheng},
title = {TimeRAG: Enhancing Complex Temporal Reasoning with Search Engine Augmentation},
year = {2025},
isbn = {9798400720406},
publisher = {Association for Computing Machinery},
address = {New York, NY, USA},
url = {https://doi.org/10.1145/3746252.3761425},
doi = {10.1145/3746252.3761425},
booktitle = {Proceedings of the 34th ACM International Conference on Information and Knowledge Management},
pages = {3230–3239},
numpages = {10},
location = {Seoul, Republic of Korea},
series = {CIKM '25}
}

@inproceedings{oldtime1,
author = {Dong, Anlei and Chang, Yi and Zheng, Zhaohui and Mishne, Gilad and Bai, Jing and Zhang, Ruiqiang and Buchner, Karolina and Liao, Ciya and Diaz, Fernando},
title = {Towards recency ranking in web search},
year = {2010},
isbn = {9781605588896},
publisher = {Association for Computing Machinery},
address = {New York, NY, USA},
url = {https://doi.org/10.1145/1718487.1718490},
doi = {10.1145/1718487.1718490},
booktitle = {Proceedings of the Third ACM International Conference on Web Search and Data Mining},
pages = {11–20},
numpages = {10},
location = {New York, New York, USA},
series = {WSDM '10}
}

@inproceedings{oldtime2, series={CIKM ’11},
   title={Recency ranking by diversification of result set},
   url={http://dx.doi.org/10.1145/2063576.2063862},
   DOI={10.1145/2063576.2063862},
   booktitle={Proceedings of the 20th ACM international conference on Information and knowledge management},
   publisher={ACM},
   author={Styskin, Andrey and Romanenko, Fedor and Vorobyev, Fedor and Serdyukov, Pavel},
   year={2011},
   month=oct, pages={1949–1952},
   collection={CIKM ’11} }

@inproceedings{oldtime3,
author = {Dong, Anlei and Zhang, Ruiqiang and Kolari, Pranam and Bai, Jing and Diaz, Fernando and Chang, Yi and Zheng, Zhaohui and Zha, Hongyuan},
title = {Time is of the essence: improving recency ranking using Twitter data},
year = {2010},
isbn = {9781605587998},
publisher = {Association for Computing Machinery},
address = {New York, NY, USA},
url = {https://doi.org/10.1145/1772690.1772725},
doi = {10.1145/1772690.1772725},
booktitle = {Proceedings of the 19th International Conference on World Wide Web},
pages = {331–340},
numpages = {10},
location = {Raleigh, North Carolina, USA},
series = {WWW '10}
}

@inproceedings{llmragtuprompt,
    title = "Narrative-of-Thought: Improving Temporal Reasoning of Large Language Models via Recounted Narratives",
    author = "Zhang, Xinliang Frederick  and
      Beauchamp, Nick  and
      Wang, Lu",
    editor = "Al-Onaizan, Yaser  and
      Bansal, Mohit  and
      Chen, Yun-Nung",
    booktitle = "Findings of the Association for Computational Linguistics: EMNLP 2024",
    month = nov,
    year = "2024",
    address = "Miami, Florida, USA",
    publisher = "Association for Computational Linguistics",
    url = "https://aclanthology.org/2024.findings-emnlp.963/",
    doi = "10.18653/v1/2024.findings-emnlp.963",
    pages = "16507--16530"
}

@article{llmirsurvey1,
   title={Large Language Models for Information Retrieval: A Survey},
   volume={44},
   ISSN={1558-2868},
   url={http://dx.doi.org/10.1145/3748304},
   DOI={10.1145/3748304},
   number={1},
   journal={ACM Transactions on Information Systems},
   publisher={Association for Computing Machinery (ACM)},
   author={Zhu, Yutao and Yuan, Huaying and Wang, Shuting and Liu, Jiongnan and Liu, Wenhan and Deng, Chenlong and Chen, Haonan and Liu, Zheng and Dou, Zhicheng and Wen, Ji-Rong},
   year={2025},
   month=nov, pages={1–54} }

@misc{cot_origin,
      title={Chain-of-Thought Prompting Elicits Reasoning in Large Language Models}, 
      author={Jason Wei and Xuezhi Wang and Dale Schuurmans and Maarten Bosma and Brian Ichter and Fei Xia and Ed Chi and Quoc Le and Denny Zhou},
      year={2023},
      eprint={2201.11903},
      archivePrefix={arXiv},
      primaryClass={cs.CL},
      url={https://arxiv.org/abs/2201.11903}, 
}

@misc{fb-cot,
      title={LightThinker: Thinking Step-by-Step Compression}, 
      author={Jintian Zhang and Yuqi Zhu and Mengshu Sun and Yujie Luo and Shuofei Qiao and Lun Du and Da Zheng and Huajun Chen and Ningyu Zhang},
      year={2025},
      eprint={2502.15589},
      archivePrefix={arXiv},
      primaryClass={cs.CL},
      url={https://arxiv.org/abs/2502.15589}, 
}

@misc{fb-cot1,
      title={Large Language Models are Contrastive Reasoners}, 
      author={Liang Yao},
      year={2025},
      eprint={2403.08211},
      archivePrefix={arXiv},
      primaryClass={cs.CL},
      url={https://arxiv.org/abs/2403.08211}, 
}

@misc{longcontext,
      title={LIFT: A Novel Framework for Enhancing Long-Context Understanding of LLMs via Long Input Fine-Tuning}, 
      author={Yansheng Mao and Yufei Xu and Jiaqi Li and Fanxu Meng and Haotong Yang and Zilong Zheng and Xiyuan Wang and Muhan Zhang},
      year={2026},
      eprint={2502.14644},
      archivePrefix={arXiv},
      primaryClass={cs.CL},
      url={https://arxiv.org/abs/2502.14644}, 
}

@inproceedings{halluci1, series={SIGIR ’25},
   title={Alleviating LLM-based Generative Retrieval Hallucination in Alipay Search},
   url={http://dx.doi.org/10.1145/3726302.3731951},
   DOI={10.1145/3726302.3731951},
   booktitle={Proceedings of the 48th International ACM SIGIR Conference on Research and Development in Information Retrieval},
   publisher={ACM},
   author={Shen, Yedan and Wu, Kaixin and Ding, Yuechen and Wen, Jingyuan and Liu, Hong and Zhong, Mingjie and Lin, Zhouhan and Xu, Jia and Mo, Linjian},
   year={2025},
   month=jul, pages={4294–4298},
   collection={SIGIR ’25} }

\end{document}